\documentclass[lettersize,journal]{IEEEtran}
\usepackage{amsmath,amsfonts}
\usepackage{algorithmic}
\usepackage{algorithm}
\usepackage{array}
\usepackage[caption=false,font=normalsize,labelfont=sf,textfont=sf]{subfig}
\usepackage{textcomp}
\usepackage{stfloats}
\usepackage{url}
\usepackage{verbatim}
\usepackage{graphicx}
\usepackage{cite}
\usepackage{amssymb}
\usepackage{booktabs}
\usepackage[utf8]{inputenc}
\graphicspath{{figures/}}
\begin{document}

\title{Capturing Electrical Windows Through Reconfigurable \\ Optical Networks: Transfer-aware Scheduling of \\ Multi-Region AI Data Centers}

\author{%
Xi Tong,
Zijian Li,
Xiaoyi Fan,~\IEEEmembership{Member,~IEEE},
Cong Zhang,~\IEEEmembership{Member,~IEEE},
Ershun Du\textsuperscript{*},~\IEEEmembership{Member,~IEEE},
Qin~Wang\textsuperscript{*},~\IEEEmembership{Senior Member,~IEEE}%

\thanks{Corresponding authors: Ershun Du and Qin Wang.}%
\thanks{X. Tong and Q. Wang are with the Department of Electrical
and Electronic Engineering, The Hong Kong Polytechnic University,
Hong Kong SAR, China (e-mail: polyu-eee-xi.tong@connect.polyu.hk; qin-ee.wang@polyu.edu.hk).}%
\thanks{Z. Li is with the College of Computer Science and Artificial Intelligence, Fudan University, Shanghai, China (e-mail: lizj21@m.fudan.edu.cn).}%
\thanks{X. Fan is with the Department of Electrical Engineering,
Tsinghua University, Beijing, China (e-mail: fan-xy25@mails.tsinghua.edu.cn).}%
\thanks{C. Zhang is with the Sichuan Energy Internet Research
Institute, Tsinghua University, Chengdu, China (email: zhangcong@tsinghua-eiri.org).}%
\thanks{E. Du is with the Lab of Low Carbon Energy, Tsinghua University, Beijing, China (e-mail: duershun@tsinghua.edu.cn).}%
}

\markboth{IEEE Transactions on Smart Grid}%
{Tong \MakeLowercase{\textit{et al.}}: Capturing Electrical Windows Through Reconfigurable Optical Networks}

\maketitle
\bstctlcite{BSTcontrol}

\begin{abstract}
Artificial-intelligence data centers (AIDCs) can provide spatiotemporal load flexibility by migrating training workloads in response to regional electricity prices and renewable-energy (RE) availability. However, transfer-blind scheduling may overestimate migration benefits by overlooking transfer time or prevent training from resuming within renewable-rich or low-price windows when considering transfer time. This paper proposed a transfer-aware compute–electricity–network scheduling framework that coordinates training placement, checkpoint transmission, and optical circuit switching (OCS) to improve local RE utilization and reduce grid-purchase cost. A compact compute--electricity master problem generates candidate schedules using estimated transfer delays. The framework then schedules optical circuits, verifies checkpoint delivery under shared-link capacity constraints, and revises infeasible schedules using updated delay estimates. It integrates day-ahead planning, event-triggered intraday updates, and real-time dispatch. In a five AIDCs benchmark with equal completed work and network resources, the proposed method reduces the real-time grid purchase cost by 12.50\% and increases the local RE utilization by 2.724 percentage points relative to static optical operation. Sensitivity and ablation analyses indicate that OCS is particularly beneficial when renewable-rich or low-price windows shift across sites and network bottlenecks hinder timely checkpoint delivery.
\end{abstract}


\section{Introduction}

\subsection{Background and Motivation}
\begin{figure}[!t]
\centering
\includegraphics[width=\columnwidth]{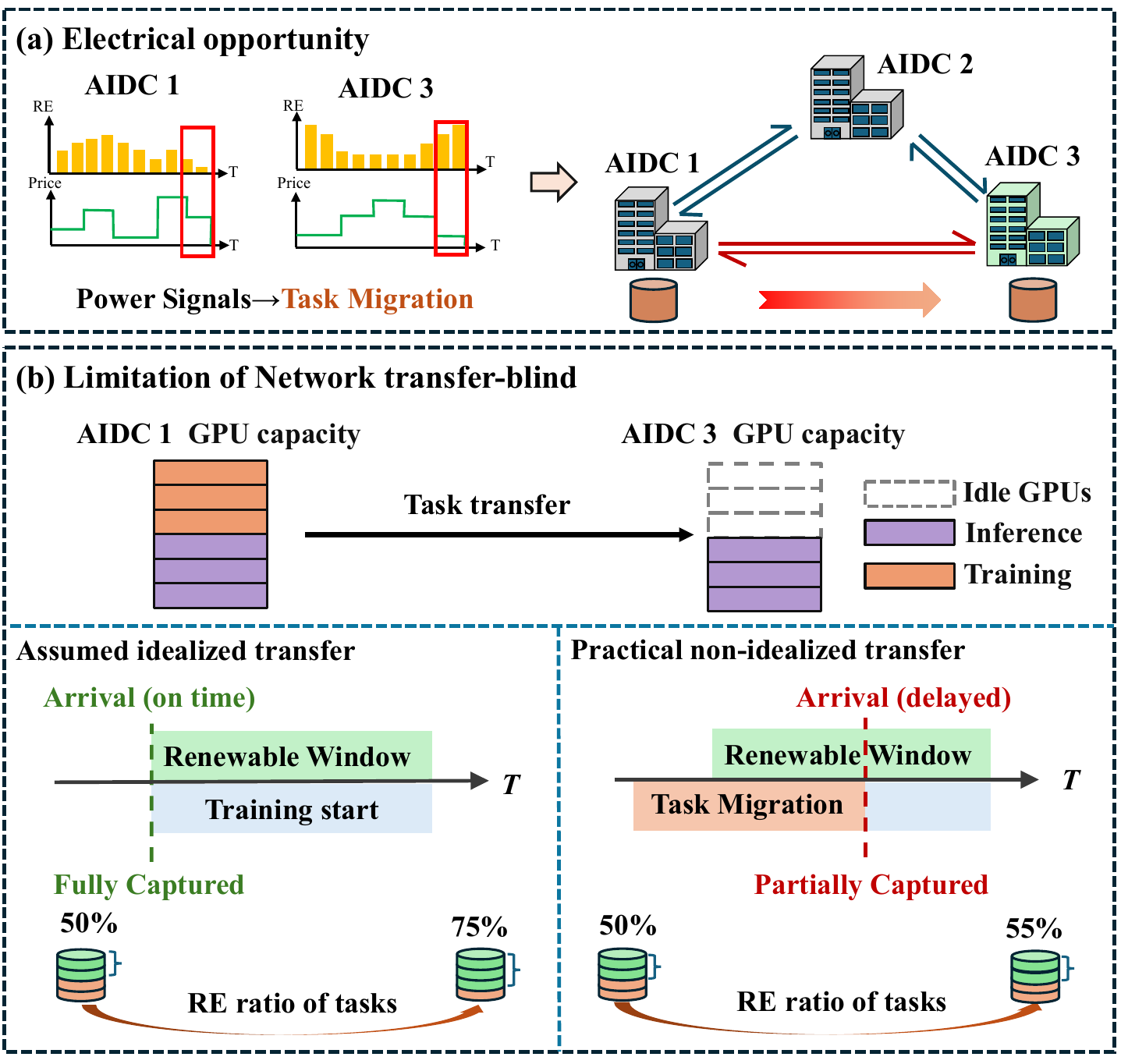}
\caption{Cross-domain migration motivation and the limitation of network transfer-blind scheduling.}
\label{fig:architecture}
\end{figure}

\IEEEPARstart{A}{rtificial-intelligence} data centers (AIDCs) are emerging as large, concentrated, and increasingly variable electric loads. Their electricity demand is rising with accelerated computing, while operational choices strongly affect the associated energy and carbon footprint \cite{IEA2024Electricity,Patterson2022Carbon}. Unlike conventional flexible resources that mainly shift consumption in time \cite{Shi2023EV,Wen2023Aggregate}, AIDCs can also relocate computing work across regions, providing both temporal and spatial flexibility \cite{Chen2021IDC,Cao2022Flex,Han2026Dispatchable,Han2026CarbonAware}. As illustrated in Fig.~\ref{fig:architecture}(a), regional differences in electricity price and renewable-energy (RE) availability therefore create strong necessity with adequate benefits for workload migration.

However, this spatial flexibility does not imply instantaneous relocation of electric load. Some power-oriented formulations abstract cross-regional workload shifting as non-physical virtual links, without representing the underlying data-delivery process \cite{Zhang2022VirtualLinks}. Other studies account for migration delay using prescribed transmission speed, distance, hop latency, or pairwise bandwidth \cite{Jin2025P2P}. Under the idealized-transfer view in Fig.~\ref{fig:architecture}(b), a successor is treated as executable once migration is selected. Explicitly accounting for data-delivery time yields the practical non-idealized view, in which late arrival can reduce the intended capture of an electrical opportunity window. The key consequence is that an electrically attractive destination may cease to be reachable before its favorable RE or price window closes. 

Therefore, the question is not only where and when training should migrate, but also whether the task transfer can finish early enough for the successor to capture the opportunity. Achieving this transfer-aware scheduling requires the underlying network to provide sufficient end-to-end bandwidth over the planned transfer interval\cite{Zhang2017Amoeba}. However, conventional networks adapt primarily through routing and rate control\cite{Jin2016Owan}, which is difficult to meet the dynamic fluctuations in demand between AIDCs. As migration demand shifts between AIDC pairs, the static networks with fixed bandwidth allocation are unable to redirect spare resources toward emerging transfer hotspots. The resulting bandwidth bottlenecks between hot AIDC pairs can delay task delivery and leave less time for workload migration within the targeted electrical windows.

\begin{figure}[!t]
\centering
\includegraphics[width=\columnwidth]{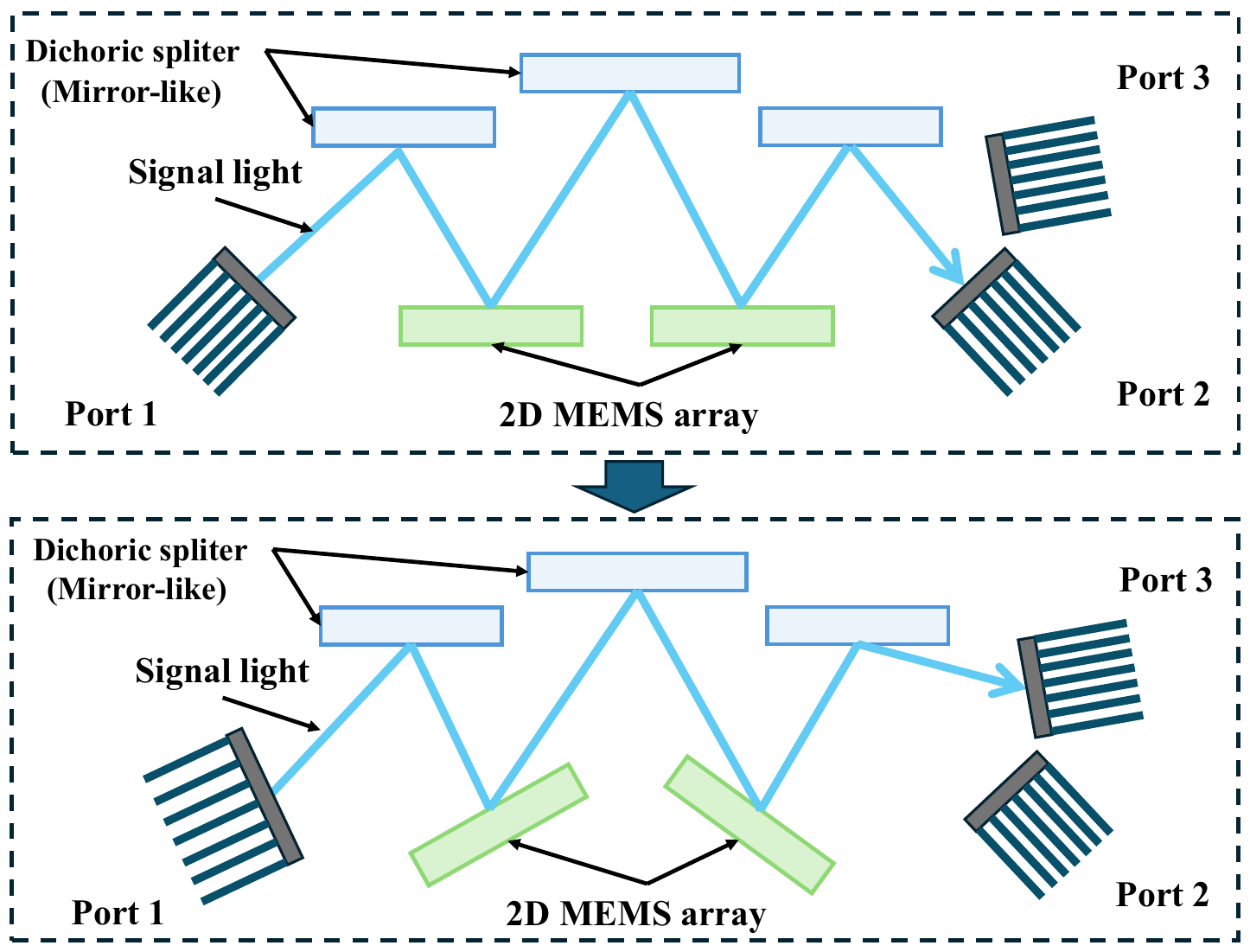}
\caption{Operating principle of a MEMS-based optical circuit switch (OCS), where software-controlled mirror steering reconfigures optical connectivity.}%
\label{fig:ocs}
\end{figure}

Optical circuit switching (OCS)  provides a practical mechanism for improving this reachability.  As illustrated in Fig.~\ref{fig:ocs}, micro electro mechanical systems (MEMS) mirrors steer optical signals between selected input and output ports, enabling millisecond-scale link/bandwidth reallocation. In practical, Google's Jupiter and Apollo demonstrate production-scale deployment of software-controlled OCS in data center networks \cite{Singh2022Jupiter,Urata2023Apollo}. OCS also requires little power to maintain mirror positions, keeping the power consumption of the optical switching layer low. Thus, OCS enables the optical bandwidth treated as a schedulable resource in compute--electricity--network coordination, adapting network support to changing migration hotspots rather than treating inter-site capacity as a fixed constraint.

\subsection{Literature Review}

Data-center flexibility research has progressed from aggregate load models to workload-aware and cross-regional formulations. Coupled regulation models coordinate computing, server, and cooling decisions \cite{Chen2021IDC}, while trace-based formulations characterize task arrivals and deadlines \cite{Cao2022Flex}. Subsequent studies have incorporated task dependencies and uncertainty when estimating deliverable flexibility \cite{Ma2024TaskDependency,Wang2026FlexQuant}. More recent multi-AIDC formulations distinguish the operating characteristics of training and inference workloads \cite{Han2026CarbonAware} and represent data size, transfer delay, and pairwise bandwidth limits \cite{Han2026Dispatchable}. These advances improve computational and temporal fidelity; however, communication capacity remains a prescribed parameter rather than a scheduling resource.

Building on these models, prior studies have quantified the power-system value of data-center flexibility. Production-scale carbon-aware operation has demonstrated temporal shifting of flexible computation toward lower-carbon periods \cite{Radovanovic2023Carbon}. Spatial scheduling has also been investigated in peer-to-peer and balancing-market settings \cite{Jin2025P2P,Cao2024NWA}, in electricity--carbon coordination \cite{Zeng2024Bilevel}, and through incentive mechanisms for power-system coordination \cite{Wu2023Incentive}. Multi-timescale and learning-based methods further improve adaptation to updated operating conditions \cite{Han2023MPC,Dvorkin2025Agent}. Collectively, these studies establish the electrical value of workload flexibility, but they do not make time-varying inter-site network capacity an endogenous resource whose allocation determines whether checkpoint transfer can finish within a favorable electrical window.

Networking research provides complementary mechanisms for managing communication scarcity. B4 and SWAN allocate paths and rates under shared-link constraints \cite{Jain2013B4,Hong2013SWAN}, while Opera introduces time-varying connectivity in reconfigurable data-center fabrics \cite{Mellette2020Opera}. Production systems such as Jupiter and Apollo demonstrate that optical circuit switching can dynamically reshape high-capacity connectivity under software control \cite{Singh2022Jupiter,Urata2023Apollo}; Lightwave Fabrics further reports at-scale reconfigurable optical fabrics for data-center and machine-learning workloads \cite{Liu2023Lightwave}. These systems primarily optimize network throughput, latency, utilization, or availability rather than the time-varying electrical value unlocked by migrated computation. Taken together, power system studies value workload placement under prescribed communication abstractions, whereas networking studies optimize communication without accounting for electrical-window reachability. 

\subsection{Research Gaps and Motivations}

Power-system studies generally treat inter-site communication as ideal or exogenous, whereas reconfigurable-network studies focus on communication performance rather than migration's time-varying
electrical value. Ignoring network constraints can overestimate
migration benefits, as checkpoint-delivery delays may prevent training from exploiting short-lived renewable-rich or low-price windows. This coupling between network-constrained migration and electrical-window capture remains insufficiently explored.

Based on these gaps, we proposed a transfer-aware compute--electricity--network scheduling framework to coordinate workload migration, network transmission, and time-varying electrical opportunities. Specifically, the main contributions are summarized as follows:
\begin{enumerate}
\item We identified a commonly overlooked network factor in the coordinated compute-electricity scheduling, termed it “transfer-blind,” and defined “electrical-window reachability” to assess the gap between current and expected transfer modes.
\item We incorporated OCS into a coupled compute--electricity--network model with checkpoint precedence of training jobs, inference-priority headroom, and network resource constraints.
\item We developed a three-stage day-ahead, event-triggered intraday, and one-minute real-time scheduling method, with transfer-lag feedback and network certification used to preserve migration executability.
\end{enumerate}

The remainder of the paper is organized as follows. Section II defines the system and evaluation framework. Section III presents the optimization model. Section IV develops the decomposition and multi-timescale coordination method. Section V reports the case study, and Section VI concludes.

\section{System Description and Scheduling Framework}
\subsection{Multi-Region AIDC and Network Description}
The research object is a portfolio of geographically separated AIDCs managed by one scheduling entity. Each AIDC is connected to a regional grid interface, observes a time-varying electricity price, and receives local RE generation. This physical supply arrangement is consistent with prior green data center systems \cite{Goiri2013Parasol}. Facility demand comprises non-schedulable base power, latency-sensitive inference, and checkpointed training workloads. The grid supplies the residual demand not served by local RE, while curtailment denotes physically available local RE that cannot be absorbed by the AIDC load.

\begin{figure}[!t]
\centering
\includegraphics[width=\columnwidth]{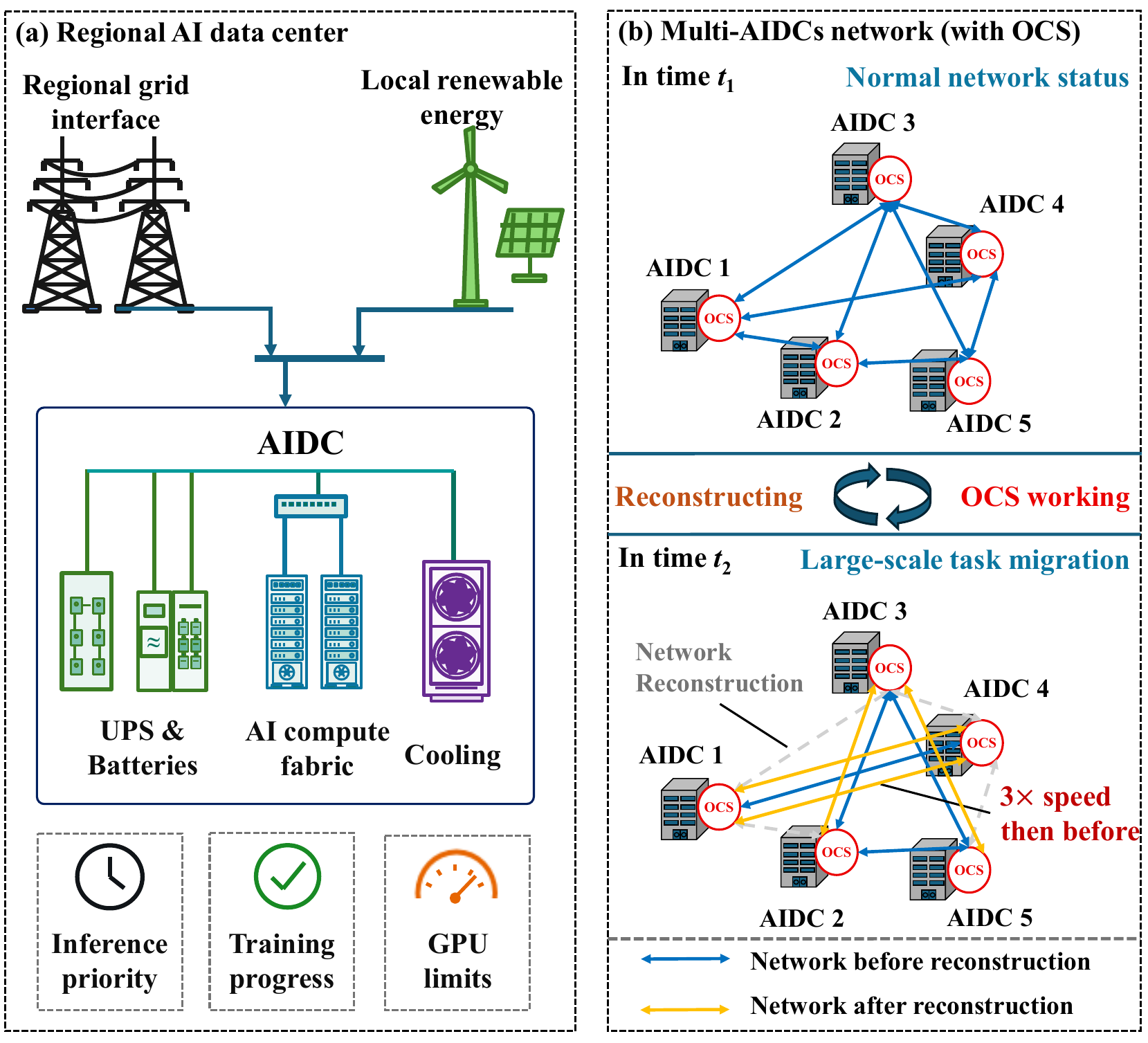}
\caption{Multi-region AIDC system and OCS network.}
\label{fig:systemandOCS}
\end{figure}

Fig.~\ref{fig:systemandOCS}(a) defines the site-level electricity--computing boundary, while Fig.~\ref{fig:systemandOCS}(b) extends it to the inter-AIDC network. Training provides temporal flexibility through delayed execution and spatial flexibility through inter-site migration; inference is served locally with priority and therefore occupies computing headroom available to training. OCS changes optical connectivity over time, whereas electricity prices and local RE availability act as exogenous scheduling signals. The battery shown in Fig.~\ref{fig:systemandOCS} is retained as physical backup infrastructure but is outside the present scheduling boundary, Cooling and other facility overheads are represented through the facility-level power conversion.

\subsection{Checkpoint Migration and Matched Network Resources}
Each training job is divided into ordered compute segments separated by checkpoints. The first segment of job must execute at its origin AIDC, only later segments may be placed at other eligible AIDCs. A site change between two consecutive segments triggers checkpoint migration, and the successor cannot start until its predecessor has completed and the checkpoint has been packed, transferred, and restored. Checkpointing provides persistent recovery states for long-running training \cite{Mohan2021CheckFreq}, while cluster schedulers demonstrate that training jobs can be preempted, migrated, and resumed across computing resources or regions \cite{Xiao2018Gandiva,Shukla2022Singularity}. In the proposed model, migration changes the spatial distribution of electric demand without reducing the required computing work.

\begin{figure}[!t]
\centering
\includegraphics[width=\columnwidth]{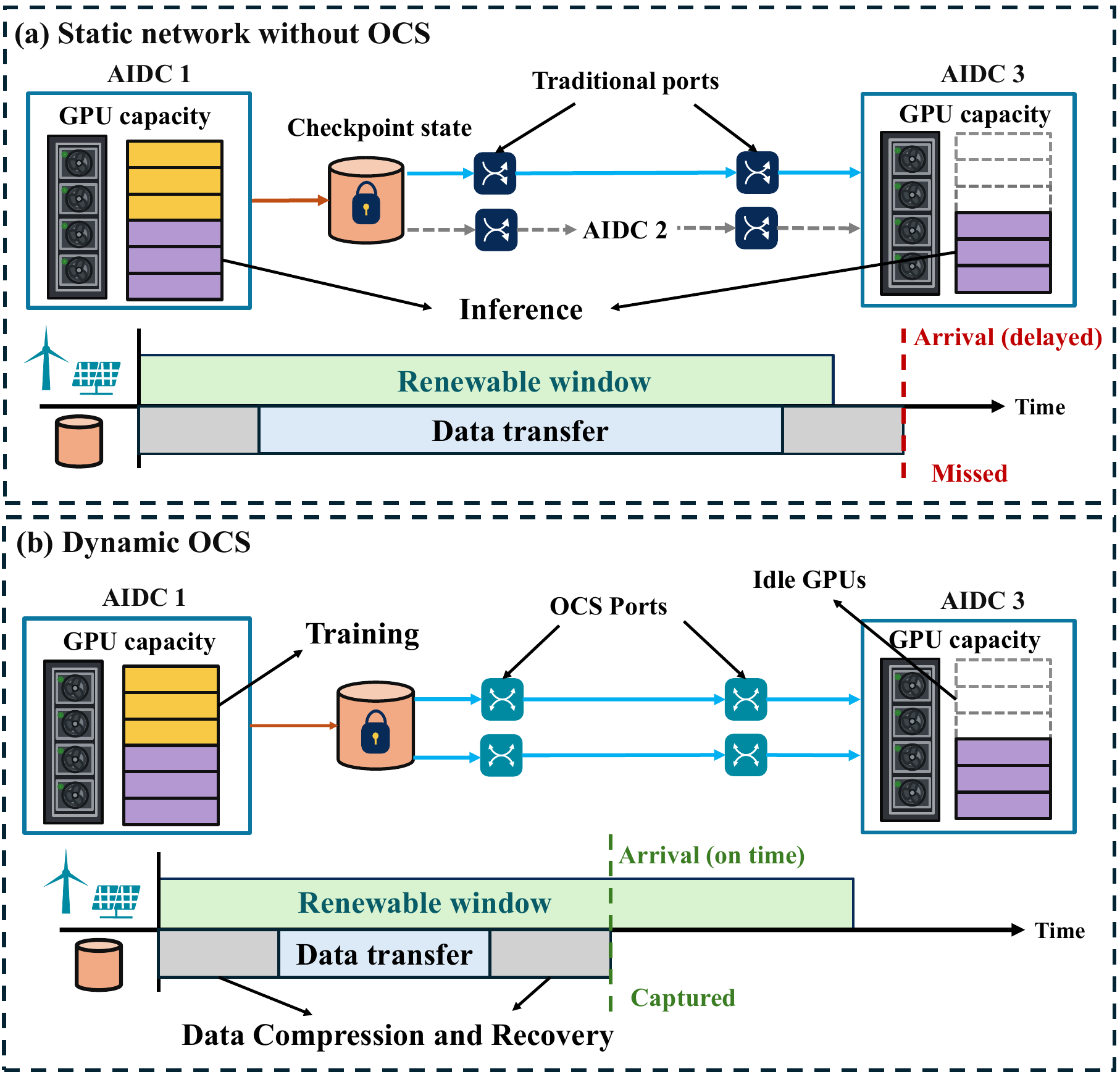}
\caption{Checkpoint-constrained RE-window reachability under matched network resources: (a) Static network without OCS and (b) Dynamic OCS.}
\label{fig:mechanism}
\end{figure}

Inter-AIDC communication uses a fixed pool of optical resources. Static optical operation keeps circuit pairings and directions fixed, whereas Dynamic OCS may re-pair or re-direct the same installed circuits as transfer demand changes. Thus, reconfiguration redistributes rather than adds communication capacity. This abstraction follows the broader principle of reconfigurable data-center networks.

We use \emph{electrical-window reachability} to describe the resulting cross-domain condition, which is a site--time interval in which local RE availability and/or electricity price makes successor execution attractive. A migration is window-reachable only if packing, network transfer, and restoration finish early enough to leave a feasible successor-execution interval before that opportunity closes. Fig.~\ref{fig:mechanism} uses an RE-rich interval as the representative electrical window. Under Static optical, limited transfer capacity can delay checkpoint completion beyond the window. Dynamic OCS can redirect existing capacity toward the binding migration and make the successor ready while the same opportunity remains available. Packing and restoration are unchanged in both cases. 

\begin{figure}[!t]
\centering
\includegraphics[width=\columnwidth]{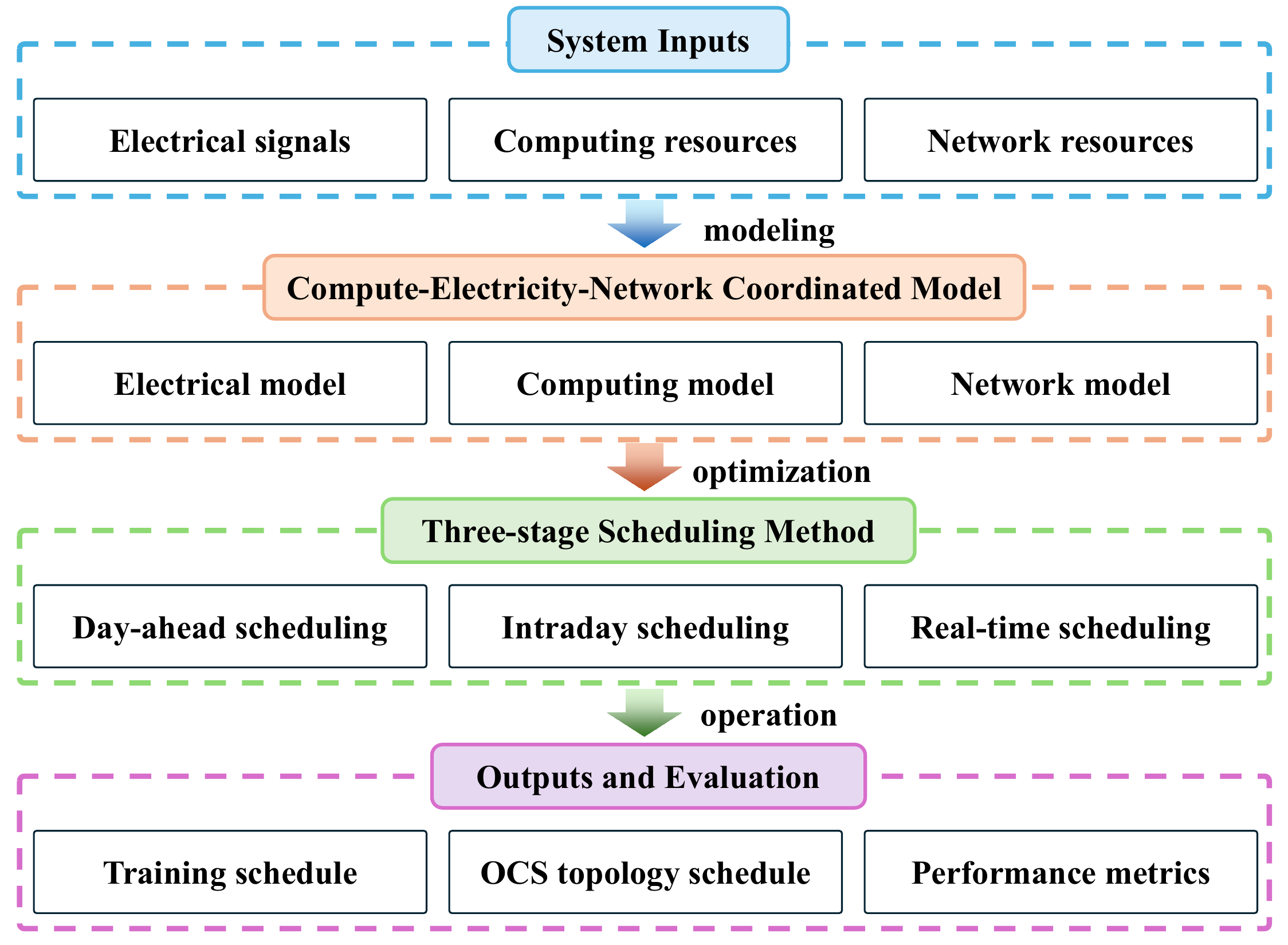}
\caption{Overall scheduling framework for transfer--aware compute--electricity--network coordination.}
\label{fig:overall_framework}
\end{figure}

\subsection{Overall Scheduling Framework}
\label{sec:overall_framework}
The preceding subsections establish the two physical foundations of the problem: geographically distributed AIDCs create time-varying electrical opportunities, while checkpoint delivery determines whether these opportunities are reachable. Meanwhile, inference demand, local RE availability, electricity price, and background traffic are updated from planning to operation. These couplings motivate the scheduling framework in Fig.~\ref{fig:overall_framework}. The figure is a schematic overview of the solution flow: it separates physical feasibility from staged information updates while retaining their coupling through explicit successor-readiness timing decisions.

At the modeling level, electrical signals, computing workloads and resources, and network resources jointly determine feasible execution, with successor readiness linking checkpoint delivery to destination load. At the operational level, day-ahead (DA) scheduling establishes the baseline plan, intraday (ID) scheduling revises uncommitted decisions when updated information becomes material, and real-time (RT) scheduling executes the latest feasible plan at one-minute resolution. Section~III formulates the coordinated model, and Section~IV presents the corresponding DA--ID--RT method.

\section{Compute--Electricity--Network Co-Optimization Model}
\label{sec:model_restored}

\subsection{Compute Scheduling}

Each job is divided into ordered, checkpoint-separated segments, consistent with fine-grained DNN checkpointing \cite{Mohan2021CheckFreq}. We first choose one admissible site--time start for each segment and keep the first segment at its prescribed origin:
\begin{subequations}\label{eq:compute_assignment}
\begin{align}
&\sum_{(i,t)\in\mathcal A_s}y_{ist}=1,\quad y_{ist}\in\{0,1\},
&&\forall s, \label{eq:compute_assignment_a}\\
&u_{is}=\sum_t y_{ist},
&&\forall i,s, \label{eq:compute_assignment_b}\\
&\tau_s=\sum_{i,t}t y_{ist},
&&\forall s, \label{eq:compute_assignment_c}\\
&\sum_t y_{i_j^0,s_1(j),t}=1,
&&\forall j. \label{eq:compute_assignment_d}
\end{align}
\end{subequations}
Where, $y_{ist}$ is the start decision, $\mathcal A_s$ is the admissible site--time set, $u_{is}$ is the resulting location, and $\tau_s$ is the start slot. Equation~\eqref{eq:compute_assignment_d} fixes the origin $i_j^0$ of the first segment $s_1(j)$.

We next convert segment starts into slot-level GPU occupancy before imposing site capacity:
\begin{subequations}\label{eq:compute_capacity}
\begin{align}
&L^{\rm tr}_{it}=\sum_s g_{j(s)}\sum_{\ell}\Gamma_{s\ell t}y_{is\ell},
&&\forall i,t, \label{eq:compute_capacity_a}\\
&L^{\rm tr}_{it}+L^{\rm inf}_{it}\le \bar G_i-G_i^{\rm safe},
&&\forall i,t. \label{eq:compute_capacity_b}
\end{align}
\end{subequations}
In \eqref{eq:compute_capacity_a}, $\Gamma_{s\ell t}$ indicates whether a duration-$H_s$ segment started at $\ell$ occupies $t$, and $g_{j(s)}$ is job GPU demand. Equation~\eqref{eq:compute_capacity_b} reserves $G_i^{\rm safe}$ from installed capacity $\bar G_i$ after serving non-deferrable inference $L^{\rm inf}_{it}$.

For each transition $q=(s,s^+)$, we then identify the predecessor--successor site pair and delay the successor when a checkpoint crosses sites:
\begin{subequations}\label{eq:compute_transition}
\begin{align}
&\sum_kz_{qik}=u_{is},\quad \sum_i z_{qik}=u_{ks^+},
&&\forall q,i,k, \label{eq:compute_transition_a}\\
&m_q=1-\sum_i z_{qii},\quad z_{qik}\in\{0,1\},
&&\forall q, \label{eq:compute_transition_b}\\
&\sum_{q\in\mathcal Q_j}m_q\le \bar M_j,
&&\forall j, \label{eq:compute_transition_c}\\
&\tau_{s^+}\ge\tau_s+H_s+\sum_{i,k}\lambda_{qik}z_{qik},
&&\forall q. \label{eq:compute_transition_d}
\end{align}
\end{subequations}
Equation~\eqref{eq:compute_transition_a} couples adjacent segment locations. Equations~\eqref{eq:compute_transition_b} and \eqref{eq:compute_transition_c} define migration $m_q$ and its job limit $\bar M_j$. In \eqref{eq:compute_transition_d}, $\lambda_{qik}$ is the screened inter-site transfer lag, with $\lambda_{qii}=0$, that is certified in Section~\ref{sec:method_restored}.

\subsection{Electricity Interface}

To expose the electrical effect of a compute schedule, we map occupied GPUs to IT power and then apply the site PUE:
\begin{equation}
P^{\rm fac}_{it}=\eta_i\!\left[P_i^{\rm base}+p_i^{\rm gpu}
(L^{\rm inf}_{it}+L^{\rm tr}_{it})\right],\quad \forall i,t.
\label{eq:power_mapping_compact}
\end{equation}
Where, $P_i^{\rm base}$ is base IT power, $p_i^{\rm gpu}$ is incremental power per active GPU, and $\eta_i$ is PUE. The resulting facility demand is supplied by physical local RE and the grid:
\begin{subequations}\label{eq:power_balance_compact}
\begin{align}
&P^g_{it}+P^{\rm use}_{it}=P^{\rm fac}_{it},
&&\forall i,t, \label{eq:power_balance_a}\\
&P^{\rm cur}_{it}=P^{\rm re}_{it}-P^{\rm use}_{it},
&&\forall i,t, \label{eq:power_balance_b}\\
&0\le P^{\rm use}_{it}\le P^{\rm re}_{it},\quad P^g_{it}\ge0,
&&\forall i,t, \label{eq:power_balance_c}\\
&P^{\rm fac}_{it}\le\bar P_i^{\rm fac},
&&\forall i,t. \label{eq:power_balance_d}
\end{align}
\end{subequations}
Equations~\eqref{eq:power_balance_a}--\eqref{eq:power_balance_c} define grid purchase $P^g_{it}$, RE use $P^{\rm use}_{it}$, and curtailment $P^{\rm cur}_{it}$ from local availability $P^{\rm re}_{it}$. Equation~\eqref{eq:power_balance_d} enforces the facility limit $\bar P_i^{\rm fac}$.

Peak and ramp exposure must remain feasible under both DA and ID forecasts. We therefore combine the larger inference forecast $\widetilde L^{\rm inf}_{it}$ with the smaller RE forecast $\widetilde P^{\rm re}_{it}$ and introduce the epigraph
\begin{equation}
\widetilde P^g_{it}=\Big[\eta_i\!\left(P_i^{\rm base}+p_i^{\rm gpu}
(L^{\rm tr}_{it}+\widetilde L^{\rm inf}_{it})\right)-\widetilde P^{\rm re}_{it}\Big]_+.
\label{eq:robust_grid}
\end{equation}
Where, $\widetilde L^{\rm inf}_{it}=\max\{L^{\rm inf,DA}_{it},\allowbreak L^{\rm inf,ID}_{it}\}$ and $\widetilde P^{\rm re}_{it}=\min\{P^{\rm re,DA}_{it},\allowbreak P^{\rm re,ID}_{it}\}$. In implementation, \eqref{eq:robust_grid} is represented by two linear lower bounds and a small positive tie-break, avoiding an additional binary variable.

\subsection{Shared Network and OCS}

We schedule checkpoint volumes over paths and time, following deadline-aware WAN traffic engineering \cite{Kandula2014Tempus}. Candidate paths have at most three hops in this study. Each checkpoint must arrive within its compute-created transfer window while concurrent flows share link capacity:
\begin{subequations}\label{eq:network_flow}
\begin{align}
&\sum_{p\in\mathcal P_{i_qk_q}}\sum_{t\in\mathcal W_q(y)}x_{qpt}=B_q,
&&\forall q, \label{eq:network_flow_a}\\
&\sum_q\sum_{p\ni e}x_{qpt}\le \widehat C_{et},
&&\forall e,t. \label{eq:network_flow_b}
\end{align}
\end{subequations}
Where, $B_q$ is checkpoint size, $x_{qpt}$ is volume sent on path $p$, and $\mathcal W_q(y)=\{\tau_s+H_s,\ldots,\tau_{s^+}-1\}$ is the available window. Equation~\eqref{eq:network_flow_a} guarantees delivery, whereas \eqref{eq:network_flow_b} couples concurrent migrations through effective capacity $\widehat C_{et}$.

Motivated by joint optical-circuit and flow control for WAN bulk transfers \cite{Jin2016Owan}, we combine residual packet and active optical capacity. For packet capacity $C_e^{\rm pkt}$, background traffic $b_{et}$, optical rate $C^{\rm ocs}$, and reconfiguration-loss fraction $\phi=t^{\rm loss}/\Delta t$,
\begin{equation}
\begin{aligned}
\widehat C_{et}=\kappa\Delta t\big(&[C_e^{\rm pkt}-b_{et}]_+\\
&+C^{\rm ocs}(o_{et}-\phi w_{et})\big),
\end{aligned}
\label{eq:network_capacity}
\end{equation}
where $\kappa$ converts $(\mathrm{Gbit/s})\!\cdot\!\mathrm{h}$ to TB, $o_{et}$ indicates an optical circuit, and $w_{et}$ indicates a new activation. Topology--routing co-design is also studied in data-center networks \cite{Zhang2021Gemini}; here, OCS obeys deployment-specific limits:
\begin{subequations}\label{eq:ocs_constraints}
\begin{align}
&\sum_{e\in\mathcal O}o_{et}\le K^{\rm ocs},
&&\forall t, \label{eq:ocs_constraints_a}\\
&\sum_{e\in\delta(i)\cap\mathcal O}o_{et}\le\bar n_i,
&&\forall i,t, \label{eq:ocs_constraints_b}\\
&o_{et}-o_{e,t-1}\le w_{et}\le o_{et},\quad
w_{et}\le1-o_{e,t-1},
&&\forall e,t, \label{eq:ocs_constraints_c}\\
&\sum_{\ell=t}^{t+H^{\min}-1}o_{e\ell}\ge H^{\min}w_{et},
&&\forall e,t. \label{eq:ocs_constraints_d}
\end{align}
\end{subequations}
Equations~\eqref{eq:ocs_constraints_a} and \eqref{eq:ocs_constraints_b} impose the circuit budget $K^{\rm ocs}$ and site-port budget $\bar n_i$. Equations~\eqref{eq:ocs_constraints_c} and \eqref{eq:ocs_constraints_d} detect activation and enforce minimum hold time $H^{\min}$. Static optical fixes $o_{et}=\bar o_e$; Dynamic OCS optimizes it under the same budgets.

\subsection{Objective Function and Constraints}

The objective penalizes energy use and undesirable grid behavior. To measure the latter under both forecast vintages, let $P_t^{\Sigma}=\sum_iP^g_{it}$ and $\widetilde P_t^{\Sigma}=\sum_i\widetilde P^g_{it}$, and define
\begin{subequations}\label{eq:grid_epigraphs}
\begin{align}
&P^{\rm pk}\ge P_t^{\Sigma},\quad P^{\rm pk}\ge\widetilde P_t^{\Sigma},
&&\forall t, \label{eq:grid_epigraphs_a}\\
&R_t\ge\pm(P_{t+1}^{\Sigma,v}-P_t^{\Sigma,v}),
&&\forall t,\ v\in\{\mathrm{nom},\mathrm{rob}\}, \label{eq:grid_epigraphs_b}\\
&R^{\max}\ge R_t,\quad R_t\ge0,
&&\forall t. \label{eq:grid_epigraphs_c}
\end{align}
\end{subequations}
Where, $P_t^{\Sigma,\mathrm{nom}}=P_t^{\Sigma}$ and $P_t^{\Sigma,\mathrm{rob}}=\widetilde P_t^{\Sigma}$. Thus, \eqref{eq:grid_epigraphs_a} measures the worst aggregate purchase, while \eqref{eq:grid_epigraphs_b} and \eqref{eq:grid_epigraphs_c} measure total and maximum ramping. The cost components are
\begin{subequations}\label{eq:objective_blocks}
\begin{align}
&J^{\rm en}+J^{\rm cur}=\sum_{i,t}(\pi_{it}P^g_{it}
+c^{\rm cur}P^{\rm cur}_{it})\Delta t,
&& \label{eq:objective_blocks_a}\\
&J^{\rm net}=c^{\rm mig}\sum_qm_q+c^{\rm sw}\sum_{e,t}w_{et},
&& \label{eq:objective_blocks_b}\\
&J^{\rm pr}=c^{\rm pk}P^{\rm pk}+c^{\rm rp}\sum_tR_t+c^{\rm mr}R^{\max},
&& \label{eq:objective_blocks_c}\\
&J^{\rm imb}=c^{\rm imb}\sum_t
|P_t^{\Sigma,\rm ID}-P_t^{\Sigma,\rm DA}|\Delta t.
&& \label{eq:objective_blocks_d}
\end{align}
\end{subequations}
Equations~\eqref{eq:objective_blocks_a}--\eqref{eq:objective_blocks_c} price energy, curtailment, migration, switching, peak, and ramps. Equation~\eqref{eq:objective_blocks_d} penalizes ID deviation from the DA aggregate schedule; its absolute value is linearized in implementation and is zero in DA. The complete planning problem is
\begin{equation}
\begin{aligned}
\min\quad &J^{\rm en}+J^{\rm cur}+J^{\rm net}+J^{\rm pr}+J^{\rm imb}\\
\mathrm{s.t.}\quad&\eqref{eq:compute_assignment}\text{--}\eqref{eq:objective_blocks}.
\end{aligned}
\label{eq:complete_problem}
\end{equation}
All cases share the same coefficients. AC-network constraints remain outside the site-level grid-interface boundary.

\section{Three-Stage Scheduling and Network Certification}
\label{sec:method_restored}

The implementation solves \eqref{eq:complete_problem} through candidate generation and exact multicommodity-flow certification. This separation keeps the master compact while preventing a compute plan from being accepted before its checkpoint transfers are verified. The resulting DA/ID plans are network-feasible over the prescribed paths, although the decomposition does not claim global optimality.

\subsection{DA--ID Rolling Coordination}

DA establishes the baseline; ID updates the remaining horizon with refreshed forecasts, following DA--ID rolling coordination for geographically distributed data centers \cite{Yang2025Coordinated}. In our implementation, committed compute starts and optical circuits are inherited at event $r$:
\begin{subequations}\label{eq:stage_inheritance}
\begin{align}
&y^{\rm ID,r}_{ist}=\widehat y^{,r-1}_{ist},
&&t\in\mathcal T_r^{\rm com}, \label{eq:stage_inheritance_a}\\
&o^{\rm ID,r}_{et}=\widehat o^{,r-1}_{et},
&&t\in\mathcal T_r^{\rm com}. \label{eq:stage_inheritance_b}
\end{align}
\end{subequations}
Where, $\mathcal T_r^{\rm com}$ is the committed block and hats denote the previously accepted plan. To decide when new information warrants replanning, we aggregate normalized changes in inference, RE, price, and background traffic. For $\mathcal V=\{L^{\rm inf},P^{\rm re},\pi,b\}$, the event score is
\begin{equation}
\delta_t=\frac{1}{|\mathcal V|}\sum_{v\in\mathcal V}
\frac{1}{|\mathcal K_v|}\sum_{k\in\mathcal K_v}
\frac{|v^{\rm obs}_{kt}-v^{\rm DA}_{kt}|}{s_{vk}}.
\label{eq:event_score}
\end{equation}
Each coarse window retains the largest $\delta_t$, with priority given to $\delta_t\ge\epsilon^{\rm ID}$. Here, $\mathcal K_v$ contains the entries of signal $v$ and $s_{vk}$ is its normalization scale. Future observations generate only the controlled forecast-error trajectory; deployment would use streaming forecasts.

\subsection{Compact Master and Exact Certification}

The compact master contains compute and electricity decisions, while $\lambda_{qik}$ summarizes the network time required by migration $q$. Before the first solve, it is conservatively initialized from residual maximum flow:
\begin{equation}
\lambda^{(0)}_{qik}=\begin{cases}
\left\lceil B_q/(\kappa\Delta t\,\alpha F^{\rm ref}_{ik})\right\rceil,&i\ne k,\\
0,&i=k.
\end{cases}
\label{eq:initial_screening_lag}
\end{equation}
Where, $F^{\rm ref}_{ik}$ is a maximum-flow estimate under lower-quantile residual packet capacity, and $0<\alpha<1$ is a safety factor. After the master produces a compute plan, the algorithm constructs a port-feasible OCS schedule and certifies all transfers by solving
\begin{subequations}\label{eq:network_certificate}
\begin{align}
&\min_{x,\sigma\ge0}\quad
\sum_q\sigma_q+\varepsilon\sum_{q,p,t}t x_{qpt},
&& \label{eq:network_certificate_a}\\
&\sum_{p,t\in\mathcal W_q}x_{qpt}+\sigma_q=B_q,
&&\forall q, \label{eq:network_certificate_b}\\
&\sum_q\sum_{p\ni e}x_{qpt}\le\widehat C_{et},
&&\forall e,t. \label{eq:network_certificate_c}
\end{align}
\end{subequations}
In \eqref{eq:network_certificate_a}, $\sigma_q$ is undelivered checkpoint volume and small $\varepsilon$ favors earlier delivery. Equations~\eqref{eq:network_certificate_b} and \eqref{eq:network_certificate_c} enforce delivery accounting and shared-link capacity. The plan is accepted only if $\Sigma=\sum_q\sigma_q\le\epsilon^{\rm net}$.

If migration $q$ violates the certificate, its observed throughput is used to enlarge only the corresponding binding lag:
\begin{subequations}\label{eq:binding_lag_update}
\begin{align}
&\Delta\lambda_q=\min\!\left\{\Delta^{\max},
\max\!\left(1,\left\lceil\frac{\sigma_q}{V_q/h_q^{\rm obs}}\right\rceil\right)\right\},
&& \label{eq:binding_lag_update_a}\\
&\lambda^+_{q i_qk_q}=\max\{\lambda_{q i_qk_q}+1,
h_q^{\rm obs}+\Delta\lambda_q\}.
&& \label{eq:binding_lag_update_b}
\end{align}
\end{subequations}
Where, $h_q^{\rm obs}$ is the tested window and $V_q=B_q-\sigma_q$ is delivered volume. Set $\Delta\lambda_q=1$ when $h_q^{\rm obs}V_q=0$. The master is warm-started after each update, and the loop stops at the tolerance or repair limit.

\subsection{Dynamic Candidate Portfolio and RT Dispatch}

The dynamic policy does not rely on a single lag estimate. It generates one candidate from the reconfigurable-network oracle and another from static-reference lags, applies the same certificate to both, and retains the lower-cost feasible candidate:
\begin{equation}
c_D^*=\underset{\substack{c\in\{c^{\rm rec},c^{\rm sta}\}:\\
\Sigma(c)\le\epsilon^{\rm net}}}{\operatorname{arg\,min}}J(c).
\label{eq:dynamic_portfolio}
\end{equation}
Because Dynamic OCS can retain the static circuit, the certified Static plan is the fallback incumbent.

RT executes the latest certified plan at one-minute resolution. To favor transfers that threaten successor starts or job deadlines, each released checkpoint receives priority
\begin{equation}
\begin{aligned}
\omega_{qm}=1&+\frac{\beta}{\max(1,s_q^{\rm plan}-m)}\\
&+\frac{0.5\beta}{\max(1,d_{j(q)}-m-H^{\rm rem}_{qm})}
+\beta^{\rm late}\mathbf 1_{qm}^{\rm late},
\end{aligned}
\label{eq:rt_urgency_compact}
\end{equation}
where $s_q^{\rm plan}$ is the planned successor start, $d_{j(q)}$ is its deadline, $H^{\rm rem}_{qm}$ is remaining chain-compute time, and $\mathbf 1_{qm}^{\rm late}$ marks non-positive slack. Every 15 min, Dynamic OCS scores port-feasible matchings by remaining volume, urgency, path length, and the certified plan. A new circuit loses its first minute. Given the selected matching, minute-scale checkpoint flow is dispatched by
\begin{subequations}\label{eq:rt_flow_dispatch}
\begin{align}
&\max_{x\ge0}\quad\sum_q\omega_{qm}\sum_p x_{qpm},
&& \label{eq:rt_flow_dispatch_a}\\
&\sum_p x_{qpm}\le B^{\rm rem}_{qm},
&&\forall q,m, \label{eq:rt_flow_dispatch_b}\\
&\sum_q\sum_{p\ni e}x_{qpm}\le\widehat C^{\rm RT}_{em},
&&\forall e,m. \label{eq:rt_flow_dispatch_c}
\end{align}
\end{subequations}
Equation~\eqref{eq:rt_flow_dispatch_a} prioritizes urgent delivered volume; \eqref{eq:rt_flow_dispatch_b} and \eqref{eq:rt_flow_dispatch_c} limit it by remaining checkpoint data and minute-level link capacity. Inference has priority and active segments are non-preemptive. A successor requires completed compute, completed transfer, and available GPUs. The main experiments disable early starts, so RT may delay but never advance the accepted plan.

\section{Case Study}
\subsection{Scenario and Data Settings}
\begin{figure}[!t]
\centering
\includegraphics[width=\columnwidth]{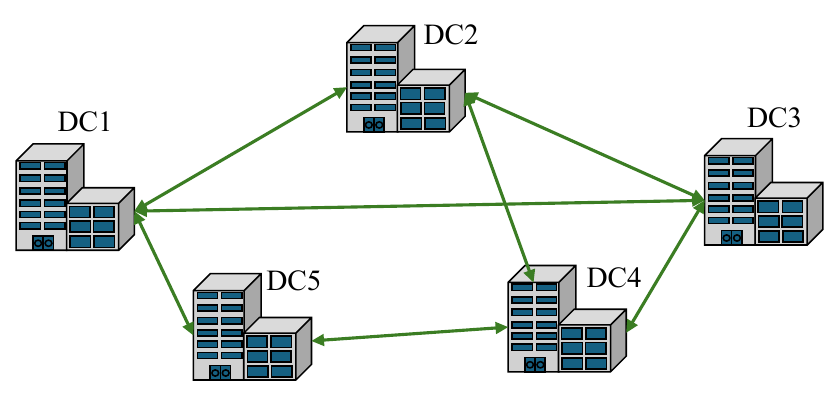}
\caption{Network connection between 5 AIDCs.}
\label{fig:AIDCs_connection}
\end{figure}

In order to test whether optical reconfiguration makes short-lived renewable-rich, low-price windows accessible to training workloads, we established five geographically distributed AIDCs, which are interconnected via a network to transmit data and information. The network connection relationship between the 5 AIDCs is shown in Fig.~\ref{fig:AIDCs_connection} . At the same time, these AIDCs are all assigned training and inference tasks of varying intensities. The inference task is considered to be highly time-latency sensitive, and thus a rigid load that cannot be transferred in either time or space. To further validate our ideas, we designed four scenarios, as showen in Table~\ref{tab:case_scenarios}. In particular, C and D both permit multi-hop routing and shared-link competition, only the circuit direction is fixed in C and reconfigurable in D.

\begin{table}[!t]
\caption{Comparison Scenarios Used in the Case Study}
\label{tab:case_scenarios}
\centering\footnotesize
\setlength{\tabcolsep}{3pt}
\begin{tabular}{@{}>{\raggedright\arraybackslash}p{0.25\columnwidth}>{\raggedright\arraybackslash}p{\dimexpr0.75\columnwidth-2\tabcolsep\relax}@{}}
\toprule
Case & Migration and network treatment \\
\midrule
A: No migration & Training remains at its assigned origin; temporal scheduling is retained. \\
B: Ideal migration & Checkpoint transfer has zero duration and no network-capacity restriction. \\
C: Static optical & Shared packet network plus one fixed 100-Gbit/s circuit, selected ex ante from a workload prior. \\
D: Dynamic OCS & The same packet network and circuit budget; the circuit can be reassigned, subject to switching loss and holding constraints. \\
\bottomrule
\end{tabular}
\end{table}

\textit{Workload data:} Training jobs originate from the Kalos trace in AcmeTrace~\cite{Hu2024Acme}. We choose 16 pretraining records with  256--1024 GPUs and 2.09--13.75 h. Durations are rounded upward to 15-min slots; arrivals are remapped into the benchmark day. The input training jobs contains assumed 2.10--12.15-TB states, the main network-pressure multiplier gives 5.775--33.413 TB per migrated state. Each job has at most four compute segments and two migrations.

Inference shapes come from successful BurstGPT requests~\cite{Wang2024BurstGPT}, aggregated by token volume into 15-min intervals. The typical day is selected by median daily token volume and the burst day by maximum interval token volume. Conversion to reserved GPUs, superposition, and site allocation are benchmark assumptions, not measured GPU utilization. The intraday inference profile blends 30\% day-ahead and 70\% realized profiles as a controlled forecast update, inference receives priority in all cases.

\textit{Electrical data and construction:} The base electricity data are from CAISO~\cite{CAISO2026OASIS}. The Daily Renewable Report supplies solar/wind forecasts and actual generation. Hourly day-ahead prices are repeated over four slots, whereas 5-min real-time prices and renewable series are averaged to 15 min after filling the two missing renewable samples per series by linear interpolation. Renewable shapes are normalized by their daily maxima and scaled to assumed AIDC capacities.

\begin{figure}[!t]
\centering
\includegraphics[width=\columnwidth]{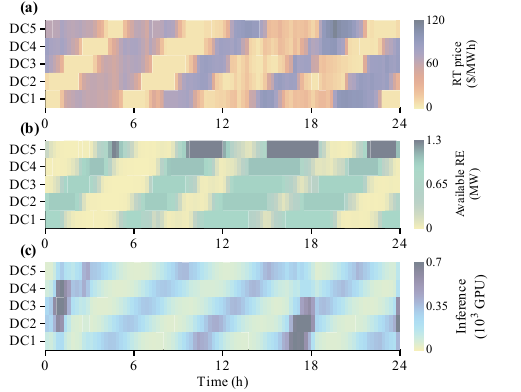}
\caption{Input conditions at all five sites (AIDC): (a) RT price, (b) available renewable power, and (c) inference GPU reservations.}
\label{fig:case_inputs}
\end{figure}

To test shifting electrical opportunities, a controlled transformation redistributes renewable availability and inference reservations across AIDCs and adds a locational price component. The transformation preserves each slot's total renewable power, total inference GPUs, and mean price. Table~\ref{tab:case_settings} gives the principal assumptions; Fig.~\ref{fig:case_inputs} displays the resulting inputs. 

\begin{table}[!t]
\caption{Main-Case Parameters and Constructed Assumptions}
\label{tab:case_settings}
\centering\footnotesize
\setlength{\tabcolsep}{2.4pt}
\begin{tabular}{@{}>{\raggedright\arraybackslash}p{0.44\columnwidth}>{\raggedright\arraybackslash}p{\dimexpr0.56\columnwidth-2\tabcolsep\relax}@{}}
\toprule
Parameter & Value \\
\midrule
GPU capacity, DC1--DC5 & 1280, 1280, 1280, 1280, 1536 \\
Base IT power (kW) & 80, 90, 80, 90, 110 \\
PUE / incremental GPU power & 1.25 / 0.7 kW \\
Renewable capacity (kW) & 900, 1000, 950, 1050, 1300 \\
Packet links / optical circuit & Seven $\times$ 100 Gbit/s / one $\times$ 100 Gbit/s \\
Price adder scale / floor & \$48/MWh / \$5/MWh \\
Curtailment / peak / ramp weights & \$25/MWh; \$0.15/kW-day; \$5/MW \\
\bottomrule
\end{tabular}
\end{table}

\subsection{Comparison of Electricity Cost and Renewable Utilization}
\begin{figure}[!t]
\centering
\includegraphics[width=\columnwidth]{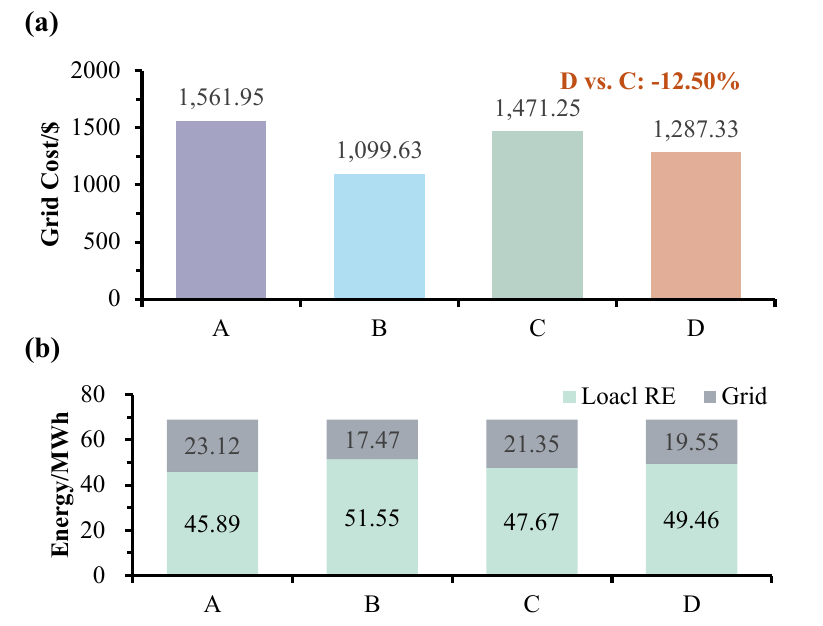}
\caption{Realized four-case comparison: (a) grid-purchase cost and (b) facility energy supplied by local RE and the grid.}
\label{fig:case_outcomes}
\end{figure}

Fig.~\ref{fig:case_outcomes} compares the economic and energy outcomes at equal completed work. All four cases consume 69.017 MWh of facility energy and finish without deadline violations, GPU overload, or incomplete segments. Relative to A's \$1561.95 cost, C saves 5.81\% and D saves 17.58\%, while B reaches \$1099.63 as the network-unconstrained reference. D therefore reduces C's cost by 12.50\% and closes 49.49\% of the observed C-to-B cost gap.

D's grid energy is 19.554 MWh, versus 21.345 MWh for C, an 8.39\% reduction. With the same 65.750 MWh of available RE, utilization rises from 72.504\% to 75.228\%, or 2.724 percentage points. This confirms actual substitution of local RE for grid purchases rather than price-only relocation. Table~\ref{tab:case_stages} shows that, from DA through ID to RT, the cost reduction remains 11.56--12.50\% and the RE-utilization gain remains 2.504--2.984 points. D schedules 20 migrations at each stage versus 12--13 for C. The sustained separation indicates that redirectable capacity preserves access to favorable sites after forecast updates and RT realization.

\begin{table}[!t]
\caption{Planning and Realized Outcomes for C and D}
\label{tab:case_stages}
\centering\footnotesize
\setlength{\tabcolsep}{3.7pt}
\begin{tabular}{llrrr}
\toprule
Stage & Case & Grid cost (\$) & RE util. (\%) & Migrations \\
\midrule
DA & C & 1492.06 & 73.536 & 12 \\
   & D & 1306.62 & 76.520 & 20 \\
ID & C & 1429.02 & 73.991 & 13 \\
   & D & 1263.79 & 76.495 & 20 \\
RT & C & 1471.25 & 72.504 & 13 \\
   & D & 1287.33 & 75.228 & 20 \\
\bottomrule
\end{tabular}
\end{table}

\subsection{Migration Scheduling and Renewable-Window Capture}
\begin{figure*}[!t]
\centering
\includegraphics[width=\textwidth]{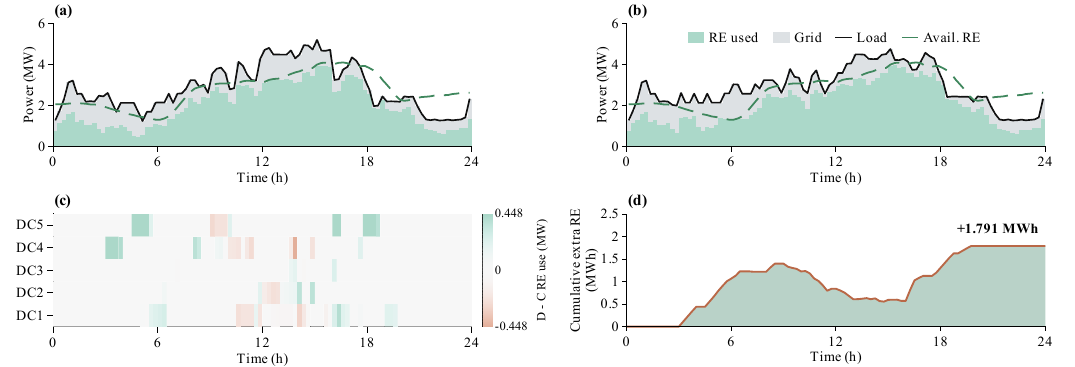}
\caption{Renewable-window capture from actual execution: (a), (b) source--load balances of C and D; (c) sitewise D-minus-C renewable use, with both gains and losses retained; and (d) cumulative extra RE absorption.}
\label{fig:case_balance}
\end{figure*}

Fig.~\ref{fig:case_balance} links the daily benefit to source--load matching. Panels (a) and (b) show that individual sites still import power when renewable availability and demand are spatially mismatched, even if aggregate renewable power is sufficient. The signed sitewise changes in panel (c) make the cumulative gain in panel (d) rise before midday, retreat during intermediate hours, and increase again after 18:00. D therefore reallocates training across a sequence of rotating site-level windows rather than increasing renewable use uniformly. The final gain is 1.791 MWh, which equal to the reduction in grid energy.

\begin{figure}[!t]
\centering
\includegraphics[width=\columnwidth]{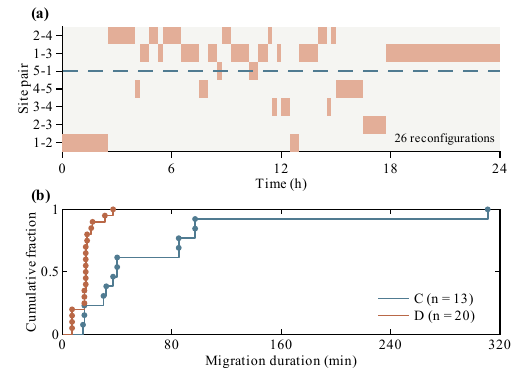}
\caption{Realized network mechanism: (a) D's active optical pair (orange) at each minute and C's fixed pair (blue dashed line); (b) empirical duration distributions of all completed migrations (C: 13; D: 20).}
\label{fig:case_network}
\end{figure}

Fig.~\ref{fig:case_network} identifies the network mechanism. C retains pair 5--1, whereas D uses all seven candidate pairs and changes the circuit 26 times. Despite executing 20 transfers versus C's 13, D reduces mean duration from 69.31 to 17.10 min (75.33\%), high-utilization exposure from 2116 to 885 edge-minutes (58.18\%), and successor-readiness lateness from 32 min to zero. Thus, the same circuit budget supports more migration decisions while delivering checkpoint state earlier. The transfer populations differ, so the duration curves are system-level rather than paired-transfer evidence; nevertheless, the concurrent reductions in duration, congestion, and lateness support the renewable-window mechanism observed in Fig.~\ref{fig:case_balance}.

\subsection{Grid-Interaction Characteristics}
\begin{figure*}[!t]
\centering
\includegraphics[width=\textwidth]{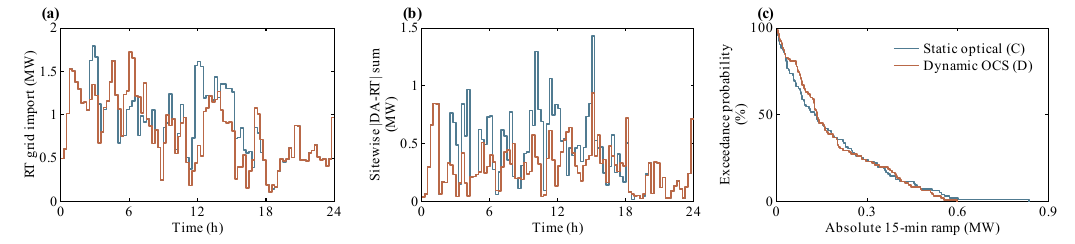}
\caption{Grid-facing trajectories: (a) aggregate RT grid import; (b) sum of sitewise absolute deviations from the corresponding DA schedule; (c) empirical exceedance of the 95 adjacent 15-min absolute ramps.}
\label{fig:case_grid}
\end{figure*}

The electrical consequences extend beyond daily energy. Fig.~\ref{fig:case_grid} and Table~\ref{tab:case_grid} show that D reduces sitewise DA--RT deviation and the maximum 15-min ramp by 29.19\% and 28.55\%, respectively. The 15-min peak decreases by 4.15\%, and the P95 ramp by 5.56\%. The stronger reductions in the maximum ramp and schedule deviation indicate that Dynamic OCS primarily limits the most consequential grid-facing corrections.

\begin{table}[!t]
\caption{Grid-Interaction Metrics and the Retained Tradeoff}
\label{tab:case_grid}
\centering\footnotesize
\setlength{\tabcolsep}{2.6pt}
\begin{tabular}{lrrr}
\toprule
Metric & C & D & Change \\
\midrule
15-min peak (MW) & 1.796 & 1.722 & $-4.15\%$ \\
Maximum ramp (MW/15 min) & 0.836 & 0.597 & $-28.55\%$ \\
P95 ramp (MW/15 min) & 0.541 & 0.511 & $-5.56\%$ \\
Sitewise DA--RT error (MWh) & 9.905 & 7.013 & $-29.19\%$ \\
Sum of absolute ramps (MW) & 17.644 & 17.834 & $+1.08\%$ \\
\bottomrule
\end{tabular}
\end{table}

Total absolute ramp nevertheless increases from 17.644 to 17.834 MW (1.08\%). Consistently, the exceedance curves remain close at lower magnitudes but separate in the upper tail. Dynamic OCS therefore reshapes rather than uniformly smooths grid imports: it lets training follow changing renewable windows while avoiding several large corrections. The grid-facing outcome is lower peak exposure, a thinner extreme-ramp tail, and closer DA schedule tracking, with slightly more cumulative ramp movement.

\subsection{Sensitivity and Ablation Analysis}
\begin{figure}[!t]
\centering
\includegraphics[width=\columnwidth]{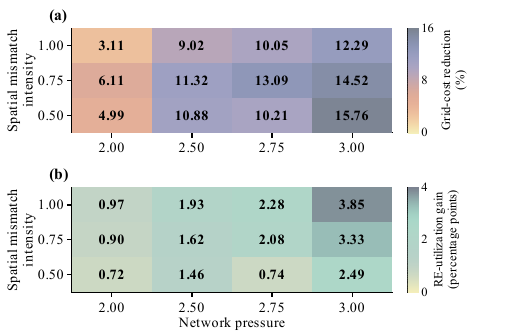}
\caption{D relative to C over the complete two-dimensional DA screen: (a) grid-cost reduction and (b) RE-utilization gain.}
\label{fig:case_sensitivity}
\end{figure}

Fig.~\ref{fig:case_sensitivity} maps the C--D difference over all 12 DA screening points. At every spatial intensity, raising network pressure from 2.0 to 3.0 increases cost reduction by 8.41--10.77 points and RE-utilization gain by 1.77--2.88 points. This repeated endpoint trend identifies transfer pressure as an amplifier of OCS value. Spatial intensity changes how that value is divided: the largest cost benefit, 15.76\%, occurs at intensity 0.5, whereas the largest RE gain, 3.85 points, occurs at intensity 1.0.

The interior cells are not monotonic. At intensity 0.5, for example, raising pressure from 2.5 to 2.75 changes the cost reduction from 10.88\% to 10.21\% and the RE gain from 1.46 to 0.74 points. This local reversal is consistent with a change in the discrete job--circuit matching regime. OCS value therefore depends on the joint alignment of transfer pressure and spatial opportunity, not on either factor independently.

\begin{table}[!t]
\caption{DA Driver Ablations: Signed D-versus-C Benefits}
\label{tab:case_ablation}
\centering\footnotesize
\setlength{\tabcolsep}{3.2pt}
\begin{tabular}{lrr}
\toprule
Drivers added to base inputs & Cost reduction (\%) & RE gain (pp) \\
\midrule
None & $-0.36$ & 0.09 \\
Price & 0.00 & 0.00 \\
Renewable redistribution & $-0.52$ & 0.08 \\
Network pressure & 4.20 & 0.71 \\
Price + renewable redistribution & 0.02 & 0.06 \\
All three & 10.05 & 2.28 \\
\bottomrule
\end{tabular}
\end{table}

Table~\ref{tab:case_ablation} isolates the joint condition. Price or renewable redistribution without added network pressure leaves the C--D differences between $-0.52\%$ and 0.02\% in cost and between 0 and 0.08 points in RE utilization. Network pressure alone yields 4.20\% and 0.71 points, while all three drivers yield 10.05\% and 2.28 points. The near-zero rows show that extra configurability alone creates no material value in this screen; clear gains emerge when electrical opportunity coincides with a constraining transfer network.

Eight of the 12 points deliver double-digit cost reductions, and five deliver RE-utilization gains above two points; no pressure-2.0 point reaches either threshold. Together with the ablation, this concentration delineates the operating region: a fixed circuit is nearly sufficient under weak transfer pressure, whereas Dynamic OCS creates clear value when the required migration direction changes across short-lived price and renewable windows. The screen thus extends the main-case mechanism to a bounded range of compute--electricity--network conditions.

\section{Conclusion}

This paper develops a transfer-aware scheduling framework that links workload placement and execution timing to checkpoint delivery and reconfigurable optical connectivity. Capturing a targeted electrical window requires checkpoint migration to finish early enough to leave a feasible successor-execution interval. Under matched optical resources, Dynamic OCS redirects capacity toward changing transfer bottlenecks, improving access to local renewable windows.

In the controlled five-AIDC benchmark, Dynamic OCS reduces RT grid-purchase cost by 12.50\% and increases RE utilization by 2.724 percentage points relative to Static optical. Sensitivity and ablation analyses indicate that the gains depend on the alignment of spatial electrical opportunities with network bottlenecks. Peak grid import, maximum ramp, and DA--RT deviations decrease, although cumulative absolute ramping increases slightly. Thus, reconfiguration can reduce large grid-import adjustments without uniformly smoothing the entire trajectory.

These findings are bounded by constructed checkpoint sizes, transformed site profiles, and a site-level grid-interface model without AC-network or reserve-market constraints. Future work should test the framework using operator measurements and examine power-flow constraints, reserve participation, and larger portfolios with multiple reconfigurable circuits.

\bibliographystyle{IEEEtran}
\bibliography{references}

\end{document}